\documentclass[10pt,conference]{IEEEtran}
\IEEEoverridecommandlockouts

\usepackage{cite}
\usepackage{graphicx}
\usepackage{float}
\usepackage{multirow}
\usepackage{amsmath,amssymb,amsfonts,amsthm}
\usepackage{xcolor}
\usepackage{makecell}
\usepackage{subcaption}
\usepackage[utf8]{inputenc}
\usepackage{tabu}
\usepackage{wrapfig}
\usepackage{hyperref}
\usepackage[all]{hypcap}
\usepackage[inline]{enumitem}
\usepackage{calc}
\usepackage{tabularx, ragged2e, caption, lipsum}
\usepackage{enumitem}   

\usepackage[most, breakable]{tcolorbox}
\usepackage{breqn}
\usepackage{pgfplots}
\pgfplotsset{compat=1.16}
\usepackage{caption}

\newcommand{\uppaalsmc}{\textsc{Uppaal}~SMC}

\newcommand{\vcd}{\mathcal{V}\mathcal{C}}
\newcommand{\newR}{r'_e}
\newcommand{\newsmcbound}{\sqrt{\newR{}(\theta) + \delta'} + \kappa}
\newcommand{\ml}{ML}
\newcommand{\cltshort}{CLT}
\newcommand{\pdf}{probability distribution function}
\newcommand{\iid}{independent identically distributed}
\newcommand{\esterror}{\kappa}
\newcommand{\lemmaparam}{\nu}

\newtheorem{lemma}{Lemma}

\newtheorem{theorem}{Theorem}

\graphicspath{{figures/}}

\def\BibTeX{{\rm B\kern-.05em{\sc i\kern-.025em b}\kern-.08em
  T\kern-.1667em\lower.7ex\hbox{E}\kern-.125emX}}

\usepackage{verbatim}

\tcbset{textmarker/.style={%
        enhanced,
        parbox=false,boxrule=0mm,boxsep=0mm,arc=0mm,
        outer arc=0mm,left=6mm,right=3mm,top=7pt,bottom=7pt,
        toptitle=1mm,bottomtitle=1mm,oversize}}
\newtcolorbox{noteBox}{textmarker, breakable,
    borderline west={6pt}{0pt}{green},
    colback=green!10!white}

\begin{document}

\textheight = 700pt
\topmargin = -62pt

\title{On the Impact of Applying Machine Learning in the Decision-Making of Self-Adaptive Systems}

\author{ 
\IEEEauthorblockN{Omid Gheibi} \IEEEauthorblockA{Department of Computer Science\\Katholieke Universiteit Leuven\\omid.gheibi@kuleuven.be} \and 
\IEEEauthorblockN{Danny Weyns} \IEEEauthorblockA{Department of Computer Science\\KU Leuven and Linnaues Universitty\\danny.weyns@kuleuven.be} \and 
\IEEEauthorblockN{Federico Quin} \IEEEauthorblockA{Department of Computer Science\\Katholieke Universiteit Leuven\\federico.quin@kuleuven.be} }

\maketitle

\thispagestyle{plain}
\pagestyle{plain}
\pagenumbering{gobble}

\begin{abstract}

Recently, we have been witnessing an increasing use of machine learning methods in self-adaptive systems. Machine learning methods offer a variety of use cases for supporting self-adaptation, e.g., to keep runtime models up to date, reduce large adaptation spaces, or update adaptation rules. Yet, since machine learning methods apply in essence statistical methods, they may have an impact on the decisions made by a self-adaptive system. Given the wide use of formal approaches to provide guarantees for the decisions made by self-adaptive systems, it is important to investigate the impact of applying machine learning methods when such approaches are used. In this paper, we study one particular instance that combines linear regression to reduce the adaptation space of a self-adaptive system with statistical model checking to analyze the resulting adaptation options. We use computational learning theory to determine a theoretical bound on the impact of the machine learning method on the predictions made by the verifier. We illustrate and evaluate the theoretical result using a scenario of the DeltaIoT artifact. To conclude, we look at opportunities for future research in this area. %

\end{abstract}

\section{Introduction}
Dealing with uncertainties is an essential problem in self-adaptive systems (SAS)~\cite{cheng2009software,Lemos10,978-3-319-74183-3_1,WeynsBook2020}. Uncertainties may affect the confidence of the adaptation decisions that need to guarantee the adaptation goals~\cite{Garlan2010,Ramirez2012,Esfahani2013,icpe14,Mahdavi2017,CalinescuMPW20}. Recently, we  witness a rapid increase in the application of supervised and interactive machine learning (\ml{}) methods to support self-adaptive software systems dealing with uncertainties~\cite{gheibi2021applying},
for example to reduce large adaptation spaces~\cite{jamshidi2019machine,quin2019efficient}, keep uncertainty parameters of runtime models up-to-date~\cite{chen2019all}, update adaptation rules under uncertainty~\cite{zhao2017reinforcement}, and predict resource consumption~\cite{ferroni2017marc}.

While these \ml{} methods aim at mitigating some types of uncertainties, they are likely sources of uncertainties themselves. In particular, since \ml{} methods apply in essence statistical techniques, they may have an impact on the decisions made by a self-adaptive system. 
This impact is particularly relevant when formal approaches are used for runtime analysis of adaptation options that aim at providing guarantees for the decisions made by a self-adaptive system~\cite{Calinescu2011,Moreno2015,abs-1908-11179,LemosGGGALSWBBB13,perpetual,8008800}. 

However, research has spent little attention on the impact of applying \ml{} methods on the decision-making in self-adaptive systems~\cite{perpetual,weyns2017software}. In fact, most self-adaptive approaches demonstrate the added value of \ml{} methods empirically based on training data sets and runtime experiments. Yet, a rigorous theoretical analysis of the impact of applying \ml{} methods on the decision-making of self-adaptive systems is often missing.  

Computational learning theory~\cite{129712.129746,Kearns94,vapnik2013nature} (CLT) studies the design, analysis, and complexity of \ml{} methods, in particular supervised methods. CLT offers a number of theorems that provide statistical bounds on the precision of machine learner models in the test phase 
(when the models are used to make predictions) based on their precision in the training phase (when the models are trained to generate the right output), the training data, and the capacity of the learner.

In this paper, 
we apply a number of CLT concepts and theorems to determine a theoretical bound for the impact of applying \ml{} on the decisions made by a self-adaptive system based in its verification results. We focus on a particular instance that combines one type of learner - linear regression  - with one type of verifier - statistical model checking - to support the decision-making of self-adaptation.  

The contribution of this paper is the definition of a theoretical bound on the guarantees provided by combining linear regression with statistical model checking in decision-making for self-adaptation. We evaluate the  theoretical bound empirically using a scenario of the DeltaIoT artifact~\cite{IftikharRBW017}. 

The remainder of this paper is structured as follows. In Section~\ref{sec:context}, we explain the research problem we tackle and we formulate the research question. 
Section~\ref{sec:preliminaries} introduces CLT concepts, definitions, and lemmas, providing the basis for our work. 
In Section~\ref{sec:solution}, we present the core contribution of this paper: a theorem that defines a theoretical bound for the impact of a machine learning method on the analysis results provided by a verifier. 
Section~\ref{sec:validation} evaluates the theoretical results using a scenario of the DeltaIoT artifact. 
Finally, in Section~\ref{sec: related works} we discuss related efforts, and we conclude with suggestions for future work in this area in 
Section~\ref{sec:conclusion}.  

\section{Problem Context and Research Question}
\label{sec:context}
A prominent approach to guarantee the adaptation goals in self-adaptive systems is the use of formal analysis of adaptation options at runtime. The principal idea of formal analysis is to use runtime models to predict the expected quality properties of different configurations that can be considered to adapt a system, i.e., the adaptation options, with the complete set  called the adaptation space. 
Over the years, a variety of formal techniques have been studied to perform runtime analysis in self-adaptive systems. One prominent approach is probabilistic model checking~\cite{TSE.2010.92,Moreno2015} that (in its basic form) exhaustively verifies the state space of each adaptation option, providing strong guarantees on the realisation of the adaptation goals. 
An alternative approach applies 
statistical model checking
~\cite{legay2010statistical} to analyse adaptation options, as for instance used in~\cite{weyns2016model}. Statistical model checking combines runtime simulation of stochastic models with statistical techniques. The approach is more efficient as exhaustive verification, yet the results are subject to an approximation interval with a confidence level.  

In this paper, we focus on self-adaptive systems that are based on the MAPE-K  model~\cite{kephart2003vision}, comprising four elements: Monitor, Analyse, Plan, and Execute that share a Knowledge repository. 
In particular, we focus on the \textit{adaptation problem} of ensuring an optimization goal for a self-adaptive system (e.g., ensuring that the packet loss of an IoT system is minimized).
The \textit{learning problem} is making predictions for a system property using a regressor to support analysis (e.g., predicting the packet loss of all the adaptation options to reduce the adaptation space based on a  threshold value of allowed packet loss). The \textit{verification problem} is making estimates to support the decision-making using a statistical model checker (e.g., estimating the packet loss of the reduced adaptation space to select the option with the lowest estimated packet loss).

The central idea of statistical model checking is to check the probability $p\in[0,1]$ that a hypothesized model $M$ (a runtime model here) of a stochastic system (the managed system with its environment) satisfies a property $\varphi$ (a quality goal), i.e., to check $P_{M}(\varphi)\geq p$ by performing a series of simulations on $M$. Statistical model checking applies statistical techniques on the simulation results to decide whether the system satisfies the property with some degree of accuracy and confidence. 
Concretely, statistical model checking allows computing an estimation of probability $p$ with an accuracy interval $[p-\epsilon, p+\epsilon]$ and confidence level $1-\alpha$. Uppaal-SMC~\cite{David2015} is a tool that supports statistical model checking. A probability estimation query in Uppaal-SMC is formulated as $p$\,$=$\,$Pr[bound](\varphi)$. 
For further details, we refer the reader to~\cite{Younes2004,Clarke2008,David2015}.  

Our interest here is how the application of \ml{} affects the analysis results and hence the adaptation decisions made based on these results. Hence, the research question we address is:  
\vspace{5pt}\\
\textit{
    What is the impact of applying a \ml{} method, in particular a linear regressor, on the analysis results, in particular the results obtained using statistical model checking, to support decision-making for self-adaptation?
}
\vspace{5pt}\\
\indent
As the problem of the research question is statistical in nature, we need to specify the behavior of a supervised machine learner statistically, for which we rely on \cltshort{}. 

\section{Preliminaries}
\label{sec:preliminaries}

This section explains essential concepts of supervised \ml{} derived from CLT. Supervised \ml{} consists of a training phase and a testing phase. During training, the machine learner trains a model using labeled examples, i.e., $(\text{input}, \text{target})$ pairs. During testing, the learner uses the trained model to predict the target for new input. Since the training data is known, the training error can be determined. Yet, in the testing phase, when the learner is used in the real world, the test data is not known in advance, and the target values it will predict will be subject to some error.
This raises the question: Can we define any bound on the test error of a learning model of a supervised learner based on its error over the training data?

\begin{figure}[!htb]
    \centering
    \includegraphics[width=0.6\linewidth]{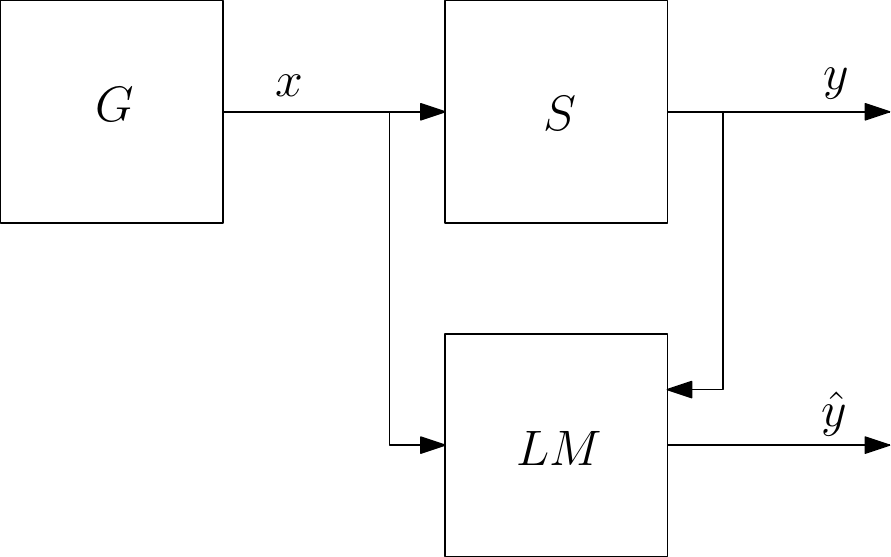}
    \caption{The model of learning~\cite{vapnik2013nature}.}
    \label{fig:model of learning}
\end{figure}

To answer this question, we formalize all the relevant concepts, leveraging on notations taken from~\cite{vapnik2013nature}. We start with devising a statistical model for a supervised learner. To that end, we use the basic model of a supervised learner shown in Figure~\ref{fig:model of learning} with the following components: 
\begin{itemize}
    \item[-]  A random generator $G$ of independent identically distributed
    vectors\footnote{Hence, we follow the common assumption that samples have been drawn from the same distribution function  independent from each other.} $x \in \mathbb{R}^n$, from a fixed and unknown probability distribution function $p(x)$. 
    \item[-] A supervisor $S$ that returns an output $y$ for every input $x$ based on a fixed and unknown conditional probability distribution function\footnote{The conditional \pdf{} is defined by $\frac{p(x,y)}{p(x)}$.
    } $p(y|x)$.
    \item[-] A learning machine $LM$ that selects a function from a set of functions $f(x;\theta)$ such that it obtains the best approximation for responses of $S$. Here, $\theta \in \Lambda$ 
    is a set of abstract parameters. Example parameters for a linear regressor $y = w \times x + b$ are $w$ and $b$.
\end{itemize}

\noindent
The function chosen by $LM$ is based on a training set $L$ that comprises of pairs  $\{(x_1, y_1), \ldots, (x_m, y_m)\}$. Note that these pairs are \iid{} and drawn from \pdf{}\footnote{Based on the definition of the conditional probability, $p(y|x) = p(x,y)/p(x)$, we have $p(x,y) = p(x)p(y|x)$.} $p(x,y) = p(x)p(y|x)$.

Suppose that the loss function\footnote{The loss function defines a distance measure on the predicted target value by the learner and the real target value.} of $y$ (the output of $S$, see Figure\,\ref{fig:model of learning}) and $\hat{y}$ (the output of $LM$) is denoted by $l(x,y,\theta) = loss(y,f(x;\theta))$, we can describe the \emph{expected risk} $r(\theta)$ as a function on the domain of $(x,y)$ pairs, denoted by $\Omega$, as:  
\[
r(\theta) = \int_{\Omega} l(x,y,\theta) p(x,y) \textbf{d}x\textbf{d}y.
\]
\noindent
The expected risk function refers to the performance of a learning algorithm in terms of the accuracy of the predictions it makes; it is defined as the continuous mathematical expectation~\cite{poznyak2009advanced} of the loss function, with $(x,y)$ being a pair of continuous random variables. 
The goal of the learning problem is to minimize the value of $r(\theta)$. However, $p(x,y)$ is unknown. Therefore, we define a replacement for $r(\theta)$ that is called \emph{empirical risk} denoted by the function $r_e(\theta)$, and defined as follows (with $\{(x_1, y_1), \ldots, (x_m, y_m)\}$ representing the training data set): 
\[
r_e(\theta) = \frac{1}{m}\sum_{i=1}^{m}l(x_i, y_i,\theta)
\]

Whereas $r(\theta)$ is based on unknown $p(x,y)$, the empirical risk corresponds to the discrete mathematical expectation of the loss function based on the known samples from $\Omega$. 
To answer the question how we can define a bound on the test error of a learning model of a supervised learner based on its error of the training data, we need to define a relation between $r_e(\theta)$ and $r(\theta)$ based on the definition of the loss function.

To that end, a vital concept is \emph{Vapnik–Chervonenkis (VC) dimension} that defines the learning capacity of the machine learner based on its loss function $l(x,y,\theta)$ -- that is a set of functions based on $\theta$ parameter -- and its input dimension, i.e., the dimension of $x$. Lemma~\ref{lem:linear vc} states that the VC dimension of linear regression methods with input dimension $n$ (we use this lemma for the evaluation case in this paper). Values for other common learning methods can be found in the \cltshort{} literature, see for instance~\cite{129712.129746,Kearns94}.
For further details on the formal definition of the VC dimension, we refer the reader to~\cite{vapnik2013nature}.

\begin{lemma}[\hspace{-0.1pt}\cite{burges1998tutorial}]
\label{lem:linear vc}
The VC dimension of a set of linear functions in $\mathbb{R}^n$ (such as for linear regression) is $n+1$.
\end{lemma}

We can now define a bound on the probability of the expected risk in a range based on: (i) the number of training data $m$, (ii) the empirical risk $r_e(\theta)$ with $\theta\in\Lambda$, (iii) the VC dimension of the learner $d = \vcd{}(l(x,y,\theta))$, (iv) the minimum and maximum value of the predefined loss function $a$ and $b$ respectively, and (v) a free variable $\eta \in [0,1]$ that is controllable by the designer, using the following lemma: 

\begin{lemma}[\hspace{-0.1pt}\cite{vapnik2013nature}]
For all functions in a set of totally bounded functions $a \leqslant l(x,y,\theta) \leqslant b$ with $\theta\in\Lambda$ and VC dimension 
$d = \vcd{}(l(x,y,\theta))$, 
for $m$ \iid{} training data with $d < m$, the probability $\mathbb{P}$ of the expected risk $r(\theta)$ between an upper and lower bound is bounded as:
\[
\mathbb{P}\left[r_e(\theta) - \delta \leqslant r(\theta) \leqslant r_e(\theta) + \delta \right] \geqslant 1-\eta
\]
\[
\delta = (b-a)\sqrt{\lemmaparam} \text{ and }  
\lemmaparam =\frac{d(\ln{\frac{2m}{d}}+1)-\ln{\frac{\eta}{4}}}{m}.
\]
\label{lem:vc bound}
\end{lemma}\vspace{-15pt}

\noindent 
Lemma~\ref{lem:vc bound} defines a lower and upper bound for the expected risk of a learner.
$\eta$ is a free variable set by the designer to balance the lower and upper bound of $r(\theta)$ and the minimum probability that the (lower and upper) bound contains the true value of $r(\theta)$.
The probability of the (lower and upper) bound for the expected risk can be set in different ways. 
For instance, we can have a tighter bound by 
decreasing the VC dimension using a different learning model. 
If we want to reach a high probability for the (lower and upper) bound by lowering $\eta$ but without changing the (lower and upper) bound (as $\lemmaparam$ contains $\eta$) and keeping the same learning model, we can for instance increase the number of training samples for the learner. 

\section{Impact of \ml{} On Verification Results}
\label{sec:solution} 

We now determine the impact of applying a ML method (regression) on the verification results (obtained through statistical model checking), answering the research question. 
Concretely, regression is used to predict the value of a quality property of all the adaptation options for a given adaptation goal. These predictions are used to reduce the adaptation options. These options are then verified using statistical model checking and the best adaptation option is selected using the estimated values of the quality property of the verified options.

The aim of statistical model checking is to estimate the true mean of the value $y$ of a quality. Yet, this estimation, denoted by $y'$, has an error that we denote by $\esterror$. The probability that $y$ lays in the $\esterror$-neighborhood of $y'$ is determined by the confidence level $1-\alpha$ of statistical model checker,\footnote{Probability estimation computes  approximation interval $p\pm\epsilon$ with confidence $1-\alpha$. If this estimation is repeated $N$ times, then the estimated interval contains the true probability $p$ at least $(1-\alpha)N$ times  for $N\rightarrow\infty$\,\cite{David2015}.} i.e.:   
\vspace{5pt}\\
\mbox{\hspace{50pt}}
  $\mathbb{P}[y'-\kappa \leqslant y \leqslant y'+\kappa] \geqslant 1-\alpha$
\vspace{-5pt}\\

\noindent
Note that error $\esterror$ is domain-dependent and can in general be expressed as a function $g(\epsilon)$, with $\epsilon$ representing the bounds of the approximation interval of a probability estimation $p\pm\epsilon$ obtained using statistical model checking. 

On the other hand, we know that the prediction of the machine learner will be determined by  $f(x;\theta)$ (defined in Section ~\ref{sec:preliminaries}) based on the input vector $x$. Hence, we need to define a loss function to measure the distance of the predicted quality value $\hat{y}$ and the true value $y$. For the purpose of our further analysis, we define the loss function as the squared error $(y-f(x;\theta))^2$ for two reasons: (i) this loss function is differentiable in all points, and (ii) by taking the square root of the loss function, we can simply obtain an error on $y$ with the same unit as the estimation error $\esterror$. In contrast, choosing a more straightforward loss function such as the absolute error $|y - f(x;\theta)|$ would not be differentiable at point zero.

Applied to this loss function, the expected risk $r(\theta)$ of  $y$ and $\hat{y}$ (see Section\,\ref{sec:preliminaries} and Fig.\,\ref{fig:model of learning}) is defined as follows: 
\vspace{-5pt}\\
\mbox{\hspace{20pt}}
$$r(\theta) = \int_{\Omega}{(y-f(x;\theta))^2p(x,y)\textbf{d}x\textbf{d}y}$$
\vspace{-5pt}\\
By taking the squared root $\sqrt{r(\theta)}$, we obtain the expected error (or expected euclidean distance) of $y$ and $\hat{y}= f(x; \theta)$.
Since this error has the same unit as the estimation error $\esterror$, the expected error of a verification result obtained with statistical model checking for an adaptation option selected based on its predicted value with a linear regressor will be $\sqrt{r(\theta)} + \esterror$. 

To express this empirically and answer the research question, we follow a three-step process: (i) we determine the impact of the result of statistical model checking on the expected and empirical risk of the regressor in Lemma~\ref{lem:vc bound}, (ii) we determine the absolute error between the value of the best-selected option of the reduced and the complete adaptation space (and its relation to the expected error $\sqrt{r(\theta)}$), and (iii) we combine the results of (i) and (ii) to determine the impact of ML - in particular regression - on the verification results - in particular those obtained using statistical model checking.
 
\subsection{Impact of verification results on \ml{} predictions}
\label{sec: answering to the impact of smc on ml}
 
In Section~\ref{sec:preliminaries}, the definition of empirical risk $r_e(\theta)$ assumed that the target values $y_i$ of the training data are exactly known, determining the true target values for $x_i$.  However, when applying statistical model checking, the values of the training data are estimated by the verifier. Hence, we modify the definition of $r_e(\theta)$ by replacing $y_i$ with $y_i \pm \esterror$. The definition of expected risk $r_e'(\theta)$ that is based on training values obtained through statistical model checking is defined as:
\[
\newR{}(\theta) = \frac{1}{m}\sum_{i=1}^{m}l(x_i, y_i \pm \esterror,\theta)
\]

To find a relation between $r_e$ and $\newR{}$, we develop the Taylor series\footnote{A series that expands a function in an infinite number of terms of its derivative. Also, it can be defined for multi-variable functions based on the function's partial differentiation of its variable which is denoted by $\partial$.} of $l(x, y\pm\esterror,\theta)$ around  constant point $c = (x_0, y_0, \theta_0)$:
\begin{dmath*}
l(x,y\pm\esterror,\theta) = l(c) + \left(\frac{\partial l}{\partial x}(c) (x-x_0)+ \frac{\partial l}{\partial y}(c) (y\pm\esterror-y_0) + \frac{\partial l}{\partial \theta}(c) (\theta-\theta_0)\right) + \cdots
\end{dmath*}

\noindent
Rephrasing the elements of the series gives:

\begin{dmath*}
  l(x,y\pm\esterror,\theta) =  \biggl[l(c) + \left(\frac{\partial l}{\partial x}(c) (x-x_0)+ \frac{\partial l}{\partial y}(c) (y-y_0) + \frac{\partial l}{\partial \theta}(c) (\theta-\theta_0)\right) + \biggr.
  \text{\small Higher order of Taylor series for }l(x,y,\theta)\biggl. \biggr]\pm \frac{\partial l}{\partial y}(c)\esterror + \xi(\esterror^2)
  = l(x,y,\theta) \pm \frac{\partial l}{\partial y}(c)\esterror + \xi(\esterror^2)
\end{dmath*}

\noindent
The first part of the series (inside the square brackets) is the Taylor series of $l(x,y,\theta)$. The last term $\xi(\esterror^2)$ shows that the lowest power of $\esterror$ in the remaining part of the series is two. If we assume that error $\esterror$ is small and $\esterror^2$ is negligible, we can replace $l(x,y\pm\esterror,\theta)$ with $l(x,y,\theta) \pm \frac{\partial l}{\partial y}(c)\esterror$ in $\newR{}$:
\begin{dmath*}
\newR{}(\theta) \approx \frac{1}{m}\sum_{i=1}^{m}\left(l(x_i,y_i,\theta) \pm \frac{\partial l}{\partial y}(c)\esterror\right)
\end{dmath*}
Rewriting the right hand side gives: 
\begin{dmath*}
\newR{}(\theta) \approx \frac{1}{m}\sum_{i=1}^{m}\left(l(x_i,y_i,\theta)\right) \pm \frac{\partial l}{\partial y}(c)\esterror
\end{dmath*}
\begin{dmath*}
\newR{}(\theta) \approx r_e(\theta) \pm \frac{\partial l}{\partial y}(c)\esterror
\end{dmath*}
\noindent This formula expresses the empirical risk $\newR{}(\theta)$ with data values obtained from statistical model checking based on empirical risk with true data values $r_e(\theta)$ and an error  $\frac{\partial l}{\partial y}(c)\esterror$.

If we assume a smooth loss function, i.e., the absolute value of the derivative of $l$ in $y$ is limited, i.e., $a' \leqslant \left|\frac{\partial l}{\partial y}\right| \leqslant b'$,
we can rewrite the formula in Lemma~\ref{lem:vc bound} as follows: \vspace{5pt}

\begin{dmath}
\label{eq: new ml inequality}
\mathbb{P}\left[\newR{}(\theta) - \delta' \leqslant r(\theta) \leqslant \newR{}(\theta) + \delta' \right] \geqslant (1-\eta) (1-\alpha)
\end{dmath}
\[
\delta' = \delta + b' \esterror
\]
This formula shows the impact of statistical model checking on the machine learner's expected and empirical risks. The initial expected risk $r_e(\theta)$ of the lower and upper bounds of the risk in Lemma~\ref{lem:vc bound} is replaced by the expected risk $\newR{}(\theta)$ for data obtained through statistical model checking. The bound $\delta$ on the probability of the risk is affected by the error $\esterror$ of statistical model checking (as a function of $\epsilon$) denoted by $\delta'$.
The probability of the (lower and upper) bound is modified by multiplying the original probability $1-\eta$ with the confidence level of the result obtained from statistical model checking $1-\alpha$ (because, the true value of each $y_i$ is located in $\esterror$-neighborhood of $y_i$ with probability $1-\alpha$). The analysis in this section relies on the assumption that $l$ is differentiable in all points, see the selection of the squared error loss function. 
\subsection{Error of best option after adaptation space reduction}
\label{sec: answering to the impact of space reduction on best option selection}

The adaptation space reduction is based on the  predictions of a quality property by a regressor. This paper analyzes the case of an optimization goal that aims at minimizing a quality property\footnote{For the analysis of a goal that maximizes a quality property, see~\cite{webpage}.} 
The domain of the value of the quality property is defined between $L_q$ and $U_q$.\footnote{$L_q$ and $U_q$ are real values, i.e., $L_q \in [-\infty, +\infty)$ and $U_q \in (-\infty, +\infty]$.} 
The adaptation space reduction is based on a cut-off value $C$, with $L_q < C < U_q$, that is, all configurations whose quality value is predicted to be 
$\leq C$ are included in the reduced adaptation space. The cutoff can be defined as a specific absolute value or as a percentage of the actual adaptation options or another approach can be used. 

We denote the best-available adaption option of the complete adaptation space by $O_w$ with a true value of $B_w$. The best option of the reduced space is denoted by $O_r$ with a true value $B_r$. 
To determine the absolute error of the best-selected option, i.e., $|B_r - B_w|$, two general cases are possible.

\begin{figure}[!htb]
    \centering
    \includegraphics[width=\linewidth]{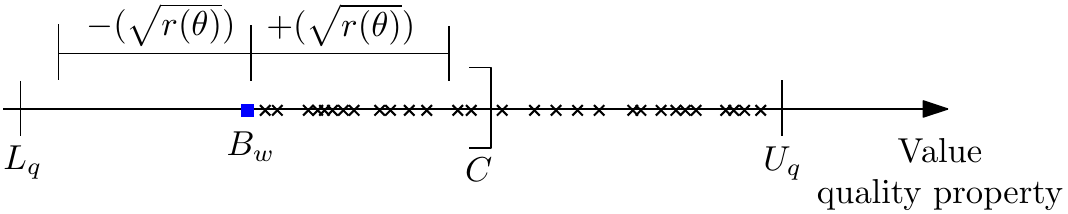}  
    \caption{The true value of the quality attribute for each adaptation option is marked with a cross on the axis. The best option of the complete adaptation space ($O_w$) is marked by the (blue) square. In this case, the predicted value of the best adaptation option with expected error $\sqrt{r(\theta)}$ is located inside the reduced space based on the cut-off value $C$ for the minimization goal. }
    \label{fig: simple reduction range}
\end{figure}

First, the prediction of the adaptation option $O_w$ with true value $B_w$ lays inside the reduced space $[L_q, C]$, i.e., $L_q \leqslant B_w -\sqrt{r(\theta)}$ and $B_w + \sqrt{r(\theta)} \leqslant C$, with $\sqrt{r(\theta)}$ the expected error of the predicted value of the selected adaptation option, as shown in Figure~\ref{fig: simple reduction range}.

\begin{figure}[!htb]
    \centering
    \includegraphics[width=\linewidth]{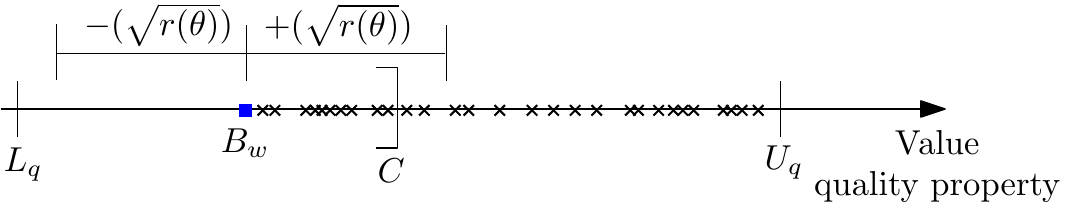}  
    \caption{The predicted value of the best adaptation option with expected error $\sqrt{r(\theta)}$ can be located outside the reduced space based on the cut-off value $C$ for the minimization goal.}
    \label{fig: complex reduction range}
\end{figure}

Second, when the prediction of $O_w$ lays outside the reduced space, its predicted value can be greater than $C$, i.e., $B_w + \sqrt{r(\theta)}>C$, see Figure~\ref{fig: complex reduction range}. 
We refer to an adaptation option with a true value in the $\sqrt{r(\theta)}$-neighborhood of $B_w$ as a \emph{feasible option}. 
Assume that the predicted values of adaptation options are \emph{uniformly distributed}\footnote{We use uniform distribution for explaining the general idea. Based on the domain and regression method, other distributions may apply.} around their true value $\sqrt{r(\theta)}$-neighborhood. Take a feasible option $O_f$ with the true value $B_f$, hence, $B_f - B_w \leqslant \sqrt{r(\theta)}$.
Based on the uniform distribution, the probability $p_f$ that the prediction of $O_f$ is located inside the reduced space is equal to the ratio of its true value  $\sqrt{r(\theta)}$-neighborhood $[B_f - \sqrt{r(\theta)}, B_f + \sqrt{r(\theta)}]$ that intersects with the reduced space $[L_q, C]$:

\begin{dmath*}
p_f = \frac{\left|\left[B_f - \sqrt{r(\theta)}, B_f + \sqrt{r(\theta)}\right] \cap \left[L_q, C\right]\right|}{\left|\left[B_f - \sqrt{r(\theta)}, B_f + \sqrt{r(\theta)}\right]\cap [L_q, U_q]\right|}
\end{dmath*}

The maximum value of the divisor of $p_f$ 
is obtained when the complete $\sqrt{r(\theta)}$-neighborhood of the true value of the feasible option is located inside the range of quality property values $[L_q, U_q]$. Hence, the maximum value for the divisor is $2\sqrt{r(\theta)}$. On the other hand, since for a feasible option $B_f - \sqrt{r(\theta)} \leqslant B_w \leqslant C$ and $ B_w \leqslant C < B_f + \sqrt{r(\theta)}$, the dividend of $p_f$ is greater than or equal to $C - B_w$. Therefore 
$p_l = C-B_w / 2\sqrt{r(\theta)} \leqslant p_f$.

\noindent
We denote the lower bound of $p_f$ with 
$p_l = C-B_w / 2\sqrt{r(\theta)}$. 
Since $p_l$ does not depend on the true value of $O_f$, the prediction of any feasible option will be located in the reduced set of adaptation options with the minimum probability $p_l$.

However, in a real application we cannot compute $p_l$ as the value $B_w$ is unknown. To address this problem, we can estimate this value by selecting the best option among the set of all predictions; we denoted the value of this best option with $\hat{B}_w$. If we assume that $\hat{B}_w$ is located inside the reduced space, as the expected error of the prediction is $\sqrt{r(\theta)}$, we can estimate the probability $p_l$ with $\hat{p}_l = \frac{C-\hat{B}_w}{2\sqrt{r(\theta)}}$.

If the prediction of any feasible option is located inside the reduced space, it is straightforward to show that the error $|B_r-B_w|$ will be at most $\sqrt{r(\theta)}$. Hence, the probability of $|B_r - B_w| \leqslant \sqrt{r(\theta)}$ is equal to the probability that at least one of the feasible options is located in the reduced space.
If we denote the number of feasible options by $n$, the probability of $|B_r - B_w| \leqslant \sqrt{r(\theta)}$ is at least $1 - (1 - \hat{p}_l)^n$, i.e., formally:
\begin{dmath}
\label{eq: best-selected options bound}
\mathbb{P}\left[|B_r - B_w| \leqslant \sqrt{r(\theta)} \right] \geqslant 1 - (1 - \hat{p}_l)^n
\end{dmath}

\begin{figure}[!htb]
    \centering
    \includegraphics[width=\linewidth]{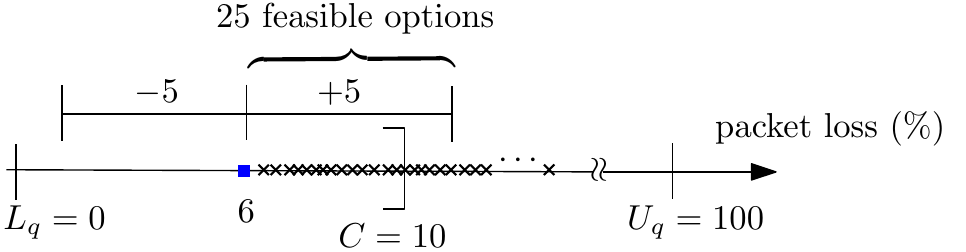}  
    \caption{Illustration error of best adaptation option after adaptation space reduction at one point in time for IoT application.}
    \label{fig:reduction-example}
\end{figure}

\textbf{Example.} 
Consider a self-adaptive IoT network with a packet loss in a range $[0, 100]$ and a cut-off value for packet loss of  $10$\,\%. The system uses a runtime model to predict the packet loss for different adaptation options.     Figure~\ref{fig:reduction-example} illustrates the error of the best adaptation option after adaptation space reduction at a particular point in time for the IoT application. 

Assume that $\sqrt{r(\theta)}$ is $5$, and at the point in time shown in the figure the best adaptation option among $n =25$ feasible options has a value $\hat{B}_w = 7$\,\%. Then, the probability that error $|B_r - B_w| \leqslant 5$ is equal to $1 - \left(1- \frac{10-7}{10}\right)^{25}\approx 0.99$. If we decrease cut-off value $C$ from $10$ to $8$, this probability, with the same number of feasible options, will decrease to $0.96$. 

\subsection{Impact of ML on Verification Results}
To answer the research question, i.e., to determine the impact of applying regression on the analysis results obtained with statistical model checking to support the adaptation decisions, we need to combine the results of the two previous sections: (i) a (lower and upper) bound of the expected risk $r(\theta)$ based on known factors (VC dimension of the learner, $\eta$, etc.) in Formula~(\ref{eq: new ml inequality}) (Section~\ref{sec: answering to the impact of smc on ml}), and (ii) 
a probabilistic bound for the absolute difference between the true value of the best-selected adaptation option in the complete and reduced adaptation space
(Section~\ref{sec: answering to the impact of space reduction on best option selection}), in Formula~(\ref{eq: best-selected options bound}). Based on that, we formulate a theorem to answer the research question:

\begin{theorem}
\label{the: main}
For the specified regressor and statistical model checker, the absolute error of the best-selected adaptation option for a minimization goal will be at most $(\newsmcbound{})$ with minimum probability $(1-\eta)(1-\alpha)^2\left(1 - \left(1 - p_m\right)^n\right)$; $\delta' = \delta + 2U_q \cdot \esterror$ 
is a measurable factor that determines the absolute difference of expected and empirical risk with $\delta = U_q^2\sqrt{\nu}$, $p_m=(C- \hat{B}_w)/(2\sqrt{\newR{}(\theta) + \delta'})$ the estimated probability that the predicted quality value of a feasible option is inside the reduced adaptation space, $\hat{B}_w$ the minimum predicted quality value of the complete adaptation options, 
and $n$ the minimum number of feasible adaptation options that can be approximated and depend on the domain at hand.
\end{theorem}

We can approximate $n$ and $p_m$ using the predictions of the regressor and relying on the verification results of the options of the reduced adaptation space.

\vspace{0.3cm}
\begin{tcolorbox}[breakable]
\textbf{Answer to Research Question:} 
The impact of applying a machine learner - a linear regressor that reduces the adaptation space with a given cutoff value for an optimization goal - on the analysis results of a verifier - a statistical model checker - is as follows: the bound on the estimation error of the best selected adaptation option increases from $\kappa=g(\epsilon)$ to $\newsmcbound{}$, while the confidence level is reduced from probability $1-\alpha$ to probability $(1-\eta)(1-\alpha)^2\left(1 - (1 - p_m)^n\right)$. $\newR{}(\theta)$ is the measured empirical risk of the regressor, $\delta'$ is the measurable factor that determines the absolute difference of the expected and empirical risk, $\eta$ is a parameter that can be set by the designer to control the tightness of the error interval and its confidence level, $p_m$ is the estimated probability that a feasible option is located in the reduced adaptation space, and $n$ is the minimum number of feasible adaptation options, that can be determined for the domain at hand.

\end{tcolorbox}

\section{Illustration and Evaluation}
\label{sec:validation}

We evaluate Theorem~\ref{the: main} empirically using a illustrative scenario of the DeltaIoT artifact\,\cite{IftikharRBW017}, inspired by the work presented in~\cite{quin2019efficient}.\footnote{For a complete description of the evaluation setting with results, see~\cite{webpage}.} 
DeltaIoT is wireless IoT network that monitors the environment and sends the data to a central facility. The network is subject to interference and varying traffic load. To that end, a managing system adjusts the communication paths to minimize packet loss. This managing system exploits a regressor that predicts the packet loss of all possible network configurations (4096 in this scenario), i.e., adaptation options, in each adaptation cycle. 
These predictions are used to reduce the adaptation space based on a cut-off value 
$C=min\,\,value + (median - min\,\,value)/4.0$. The analysis of the runtime model of the selected adaptation options is done with  \uppaalsmc{} using the verification query \texttt{Pr[<=1](<>Network.PacketLoss)} (for details see~\cite{webpage,quin2019efficient}). The minimum and the median values of $C$ are computed over the predictions.
We compare the theoretical bound on the error of the packet loss of the best selected adaptation option over 200 cycles as defined in Theorem~\ref{the: main}, i.e., $\newsmcbound{}$, and the measured error, i.e., the absolute difference between the minimum packet loss value for the optimal configuration obtained from verification of all adaptation options and the value obtained from the reduced set of adaptation options. 
Here, the empirical training risk $\newR{}$ is defined as the mean squared error (MSE) of the regressor on the training data, i.e., $\newR{}(\theta) = \frac{1}{m}\sum_{i=1}^m(y_i - \hat{y}_i)^2$.

\begin{figure}[!htb]
    \centering
    \includegraphics[width=\linewidth]{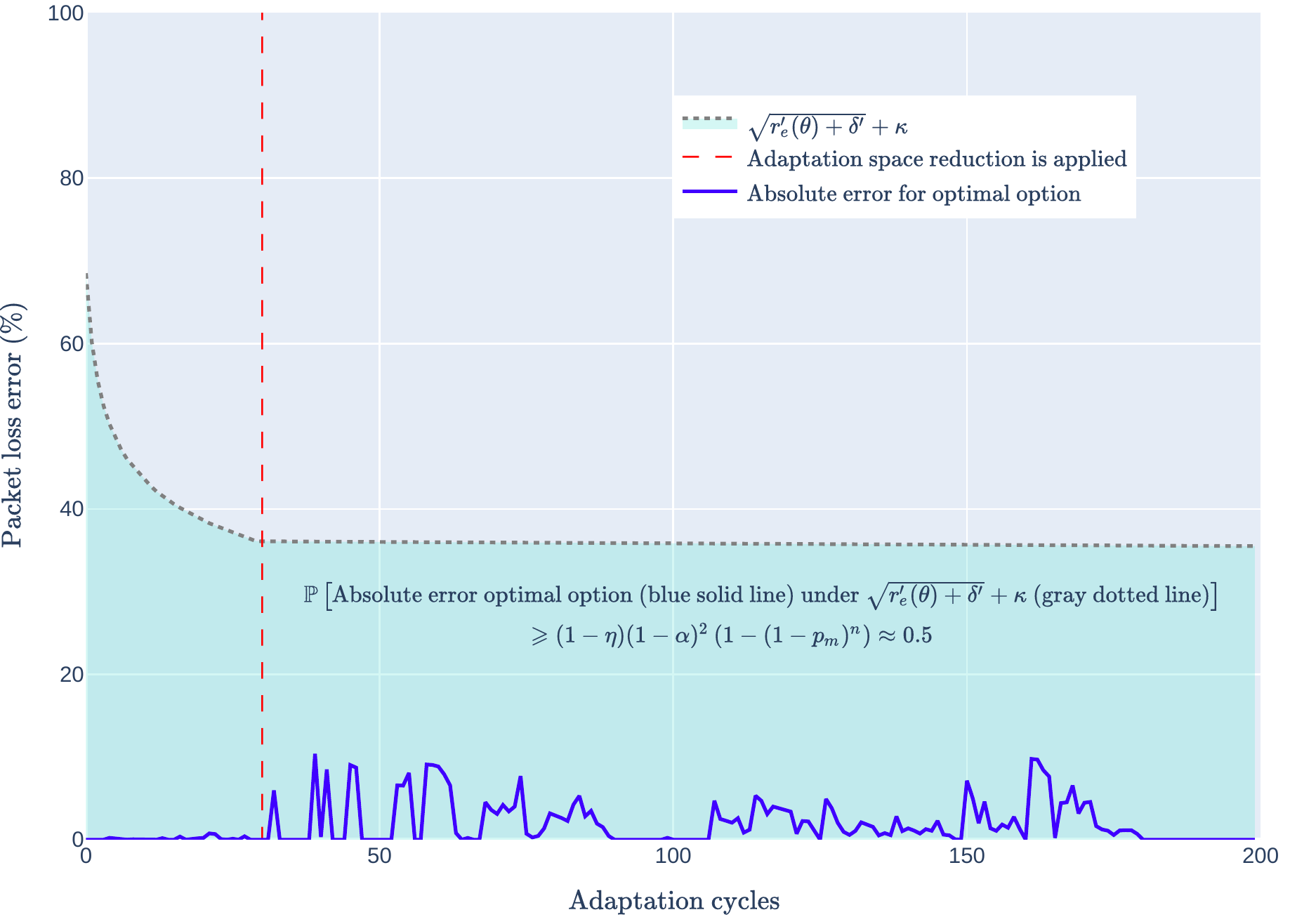}  
    \caption{The dotted line shows the theoretical bound of the  packet loss error; the solid line shows the measured error. Settings used in the experiment: 
    statistical model checking parameters $\epsilon = 0.01$ and $\alpha = 0.1$,  
    dimension of input data to determine the features of the regressor $85$, the VC dimension of the regressor $d = 85 + 1$ (see Lemma~\ref{lem:linear vc}), 
    and estimation error $\esterror = 1$, that is, $\esterror$ is equal to $g(\epsilon) = 100 \times \epsilon$.  
    %
    }
    \label{fig:new bound in each iteration}
\end{figure}

Figure~\ref{fig:new bound in each iteration} shows the results. 
During the first 30 cycles, the learner is trained without adaptation space reduction. 
From cycle 31 onward adaptation space reduction is applied. The absolute difference between the packet loss values of the best of all adaptation options and the best option of the reduced space fluctuates between 0.0 and 10.4 (mean 2.1, std. 2.7). 
The results show that the error for the packet loss obtained from this experiment comply with the theoretical bound on the estimated error and its confidence level as defined in Theorem~\ref{the: main}. 
Yet, the results are subject to validity concerns with respect to a single evaluation in one particular domain.

\section{Related Work}
\label{sec: related works}
A characteristic work that applies \ml{} in support of obtain guarantees for the adaptation goals is\,\cite{rodrigues2018learning}. That work used a decision tree to predict how long it takes to change a state in the environment based on the current state. This knowledge is then used to tune the adaptation policies. 
However, that work does not provide guarantees on the bound of the prediction results, which could be obtained leveraging on the results we present in this paper. 
Another recent study~\cite{camara2020} employs Q-learning to predict the quality of service of a system in support to the planning stage. 
While the approach can reduce the decision space effectively, the paper does not provide a theoretical bound on the guarantees for the proposed approach. 
One way to provide such guarantees is to exploit CLT that provides theoretical bounds for reinforcement learning methods, in line with the approach and results we present in this paper. 

Verification of properties of machine learners is a recent line of research applied to safety-critical systems. Reluplex~\cite{katz2017reluplex} is a characteristic work in this realm that introduced an efficient satisfiable modulo theories solver to verify a deep neural network (DNN) with a rectified linear unit activation function. 
The method was applied in the domain of collision avoidance of unmanned aircrafts. 
Other efforts related to Reluplex are~\cite{bastani2016measuring,pulina2010abstraction,ehlers2017formal}, and~\cite{bunel2018unified} that proposed a unified framework for verifying properties of DNN. Another related effort proposed a process to select safe actions based on verified results using reinforcement learning~\cite{mason2017assured}. 
In contrast to these efforts, our work studies the impact of a machine learner on the adaptation decisions made based on formal verification. 

\section{Conclusion and Outlook for Future Work}
\label{sec:conclusion} 

This paper contributes a theorem that defines a theoretical bound on the impact of applying a machine learning method, in particular a linear regressor that reduces the adaptation space of a self-adaptive system for an optimization goal, on the analysis results of a verifier, in particular a statistical model checker, that are used for making adaptation decisions. We evaluated the theorem using a simply IoT application. 

The work presented in this paper can be extended to classification tasks that apply for the decision-making of threshold goals. This case can rely on Lemma~\ref{lem:vc bound}, yet, for this case, the analysis needs to determine the impact of the classification loss on the bound of the error of the selected adaptation options and its probability. 
For \ml{}-based reduction methods in general, it may be useful to study Rademacher complexity~\cite{shalev2014understanding} to determine a tighter bound for the guarantees of verification results. Rademacher complexity
measures the learning capacity of the machine learner. Utilizing this complexity provides an alternative way to determine a bound between the expected and the empirical risk, analogous to Lemma~\ref{lem:vc bound}.
Finally, more research is required to investigate other types of learners and other use cases of ML methods used in self-adaptive systems.

\bibliography{main}
\bibliographystyle{ieeetr}
\end{document}